\documentclass[sigconf]{acmart}

\usepackage{booktabs} % For formal tables
\usepackage{multirow}
\usepackage{subfigure}
\usepackage{balance}
\usepackage{threeparttable}

\setcopyright{none}
\copyrightyear{2019} 
\acmYear{2019} 
\setcopyright{acmcopyright}
\acmConference[GLSVLSI '19]{Great Lakes Symposium on VLSI 2019}{May 9--11, 2019}{Tysons Corner, VA, USA}
\acmBooktitle{Great Lakes Symposium on VLSI 2019 (GLSVLSI '19), May 9--11, 2019, Tysons Corner, VA, USA}
\acmPrice{15.00}
\acmDOI{10.1145/3299874.3318038}
\acmISBN{978-1-4503-6252-8/19/05}
\renewcommand\footnotetextcopyrightpermission[1]{} % removes footnote with conference information in first column

\begin{document}
\title{Clockless Spin-based Look-Up Tables with Wide Read Margin}
%\titlenote{Produces the permission block, and copyright information}
%\subtitle{Extended Abstract}
%\subtitlenote{The full version of the author's guide is available as
%  \texttt{acmart.pdf} document}

\author{Soheil Salehi, Ramtin Zand, Ronald F. DeMara}
%\orcid{0000-0001-5998-8795}
\affiliation{%
  \institution{Department of Electrical and Computer Engineering, University of Central Florida, Orlando, FL, 32816 USA}
%   \streetaddress{P.O. Box 162362}
%   \city{Orlando}
%   \state{Florida}
%   \postcode{32816-2362}
}
% \email{soheil.salehi@knights.ucf.edu}

% \author{Soheil Salehi}
% \orcid{0000-0001-5998-8795}
% \affiliation{%
%   \institution{Department of Electrical and Computer Engineering\\ University of Central Florida}
%   \streetaddress{P.O. Box 162362}
%   \city{Orlando}
%   \state{Florida}
%   \postcode{32816-2362}
% }
% \email{soheil.salehi@knights.ucf.edu}

% \author{Ramtin Zand}
% \orcid{0000-0001-6864-7255}
% \affiliation{%
%   \institution{Department of Electrical and Computer Engineering\\ University of Central Florida}
%   \streetaddress{P.O. Box 162362}
%   \city{Orlando}
%   \state{Florida}
%   \postcode{32816-2362}
% }
% \email{ramtinmz@knights.ucf.edu}

% \author{Ronald F. DeMara}
% \orcid{0000-0001-6864-7255}
% \affiliation{%
%   \institution{Department of Electrical and Computer Engineering\\ University of Central Florida}
%   \streetaddress{P.O. Box 162362}
%   \city{Orlando}
%   \state{Florida}
%   \postcode{32816-2362}
% }
% \email{ronald.demara@ucf.edu}
\renewcommand{\shortauthors}{S. Salehi et al.}

\begin{abstract}
In this paper, we develop a $6$-input fracturable non-volatile Clockless LUT (C-LUT) using spin Hall effect (SHE)-based Magnetic Tunnel Junctions (MTJs) and provide a detailed comparison between the SHE-MTJ-based C-LUT and Spin Transfer Torque (STT)-MTJ-based C-LUT. The proposed C-LUT offers an attractive alternative for implementing combinational logic as well as sequential logic versus previous spin-based LUT designs in the literature. Foremost, C-LUT eliminates the sense amplifier typically employed by using a differential polarity dual MTJ design, as opposed to a static reference resistance MTJ. This realizes a much wider read margin and the Monte Carlo simulation of the proposed fracturable C-LUT indicates no read and write errors in the presence of a variety of process variations scenarios involving MOS transistors as well as MTJs. Additionally, simulation results indicate that the proposed C-LUT reduces the standby power dissipation by $5.4$-fold compared to the SRAM-based LUT. Furthermore, the proposed SHE-MTJ-based C-LUT reduces the area by $1.3$-fold and $2$-fold compared to the SRAM-based LUT and the STT-MTJ-based C-LUT, respectively. 
\end{abstract}




\maketitle
\section{Introduction}
\label{sec:introduction}
Flexibility and runtime adaptability are two of the main motivations for the wide adoption of reconfigurable fabrics. Among the most commonly used reconfigurable fabrics, Field Programmable Gate Arrays (FPGA) have been the primary focus due to their flexibility that allows realization of logic elements at medium and fine granularities while incurring low non-recurring engineering costs and rapid deployment to market. Additionally, FPGAs have been researched as promising platform that can be utilized effectively to increase reliability in case of process-voltage-temperature variation \cite{Al-Haddad2015AdaptiveFabrics}. The main challenge of static random access memory (SRAM)-based FPGAs is their increased area and power consumption to achieve flexible design. The main components of FPGAs are Look-Up Tables (LUTs) and switch boxes that are mainly consisted of SRAM cells \cite{Kuon2008FpgaChallenges}. However, SRAM-based LUTs incur limitations such as high static power, volatility, and low logic density.

Innovations using emerging devices within FPGAs have been sought to bridge the gaps needed to overcome the limitations of SRAM-based FPGAs. High-endurance non-volatile spin-based LUTs have been studied in the literature as promising alternatives to SRAM-based LUTs, Flash-based LUTs, and other state-of-the-art emerging LUTs such as resistive random access memory (RRAM)-based LUTs and phase change memory (PCM)-based LUTs \cite{Zand2017Radiation-hardenedTolerance,Tang2016AArchitectures,Huang2014AElement,Attaran2018StaticDesigns,Suzuki_2019,Suzuki2013AStructure}. Spin-based devices offer non-volatility, near-zero static power, high endurance, and high integration density \cite{Salehi2017SurveyResiliency,Yoda2017High-SpeedVoCSM}. The spin-based LUTs presented in the literature \cite{Zand2017Radiation-hardenedTolerance,Tang2016AArchitectures,Huang2014AElement,Attaran2018StaticDesigns,Suzuki_2019,Suzuki2013AStructure} require separate read and write operations as well as a clock, which makes these LUTs a suitable candidate for sequential logic operations. However, the main challenge that has not been addressed in the literature is providing a spin-based LUT design for combinational logic operation without the need for a clock. Additionally, proposed spin-based LUTs proposed in the literature fail to maintain a wide sense margin and high reliability without incurring significant area and power dissipation overheads \cite{Zand2017Radiation-hardenedTolerance,Tang2016AArchitectures,Huang2014AElement,Attaran2018StaticDesigns, Suzuki_2019,Suzuki2013AStructure}. In this paper, in order to address the aforementioned challenges, we develop a clockless $6$-input fracturable non-volatile Combinational LUT (C-LUT) with wide read margin using spin Hall effect (SHE)-based Magnetic Tunnel Junction (MTJ) and provide a detailed comparison between the SHE-MRAM and Spin Transfer Torque (STT)-MRAM C-LUTs. Additionally, we provide detailed analysis on the reliability of our proposed C-LUT in the presence of Process Variation (PV).

\section{Realizing Fracturable 6-Input Clockless LUT}
\label{sec:clut}
\begin{figure}[!t]
\centering
\subfigure[]{
\includegraphics[width=3.1in]{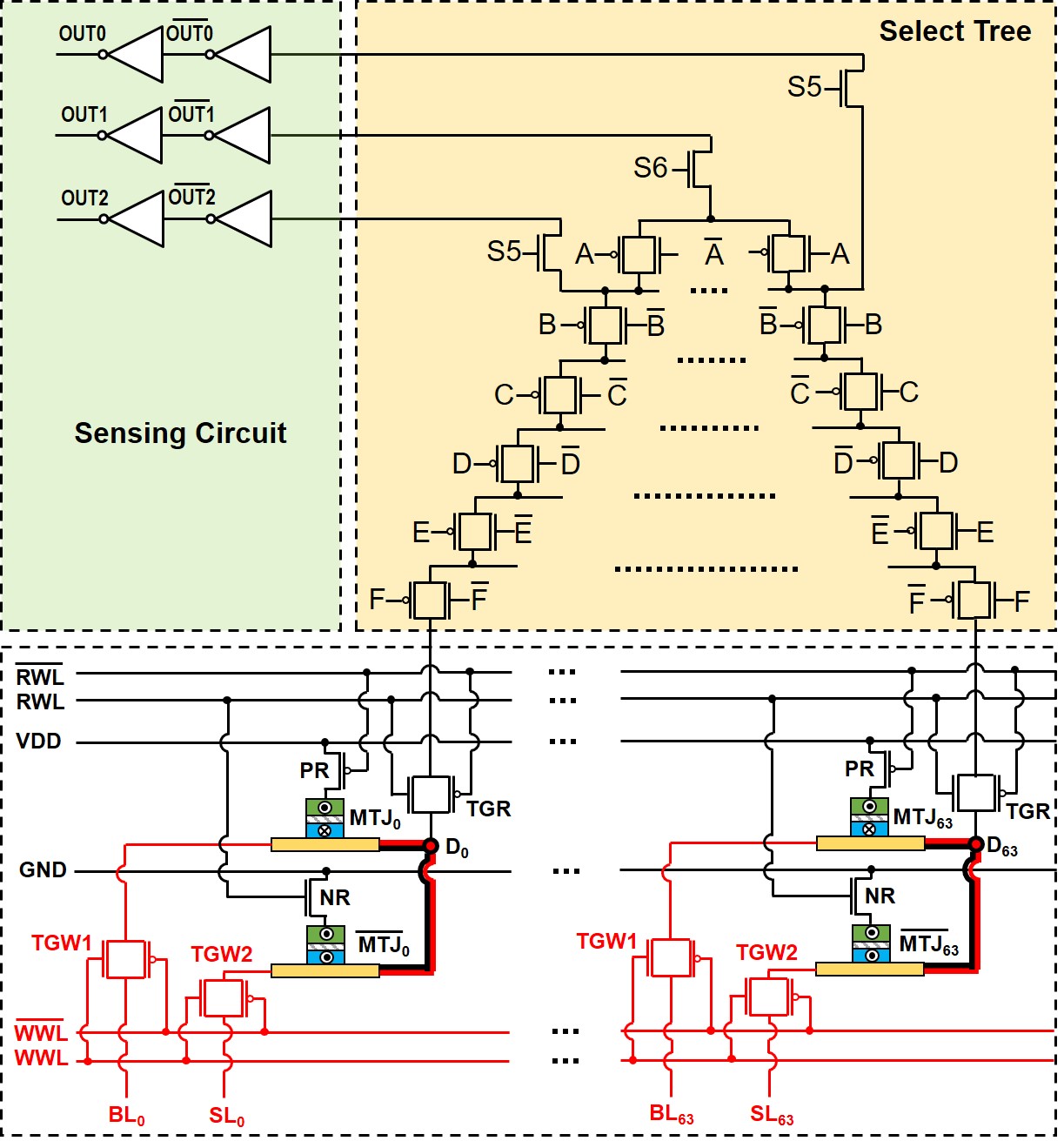}
\label{subfig:fracclut}
}
\subfigure[]{
\includegraphics[width=3.1in]{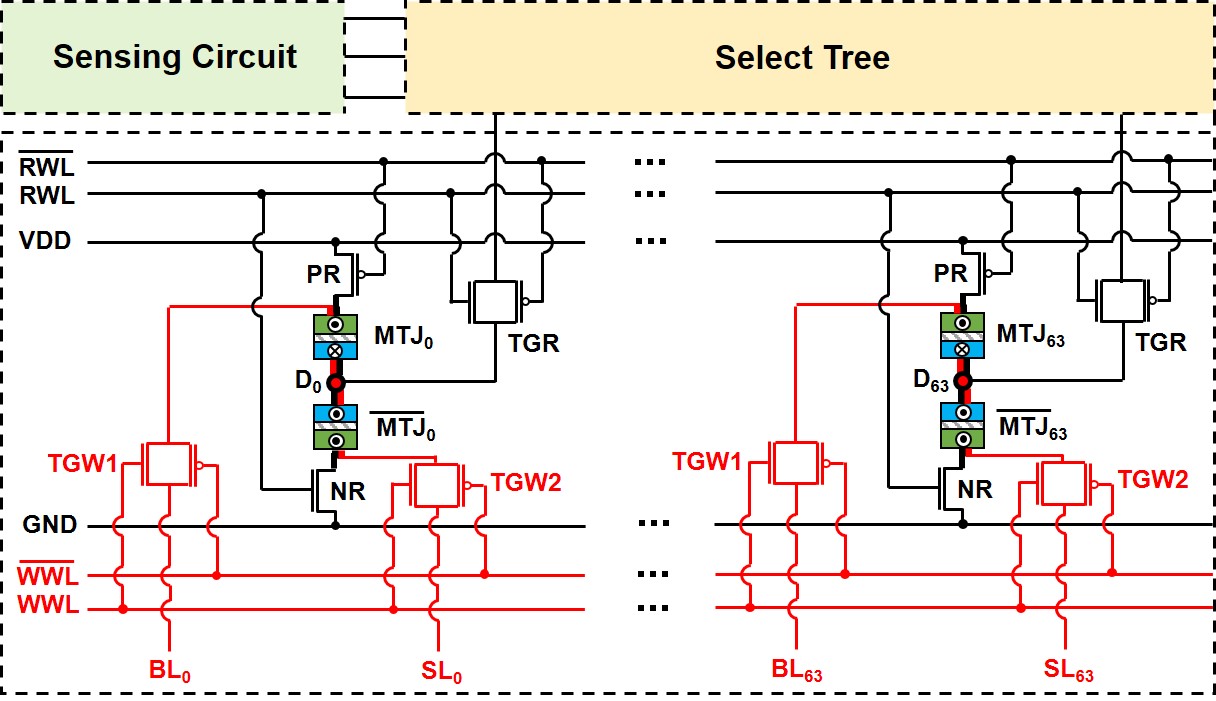}
\label{subfig:fracsttclut}
}
%\vspace{-2mm}
\caption{The circuit-level diagram of the proposed $6$-input fracturable Combinational Look-Up Table (C-LUT) using \subref{subfig:fracclut} SHE-MTJ devices and \subref{subfig:fracsttclut} STT-MTJ devices.}
\label{fig:lutfig}
\end{figure}

The primary goal of using LUTs in the reconfigurable fabrics is for implementing combinational logic. Generally, $M$-input Boolean functions are implemented using LUTs that are considered a memory that has $2^M$ memory cells. The inputs are assigned using a select tree which is constructed with Pass Transistors and Transmission Gates (TGs) \cite{Zand2016ScalableDesign}. Most contemporary FPGAs, utilize fracturable $6$-input LUTs in their design in order to be able to implement one $6$-input boolean function or two $5$-input boolean functions \cite{Percey2007AdvantagesArchitecture}. Fig. \ref{subfig:fracclut} depicts our proposed $6$-input fracturable SHE-MRAM C-LUT and Fig. \ref{subfig:fracsttclut} illustrates the $6$-input fracturable STT-MRAM C-LUT. In Fig. \ref{subfig:fracclut} and Fig. \ref{subfig:fracsttclut}, where red color indicates the write path and black color indicates the read path. When the $\textbf{WWL}$ and $\overline{\textbf{WWL}}$ signals are asserted, the Write TGs of each memory cell, $\textbf{TGW1}$ and $\textbf{TGW2}$, will turn on and using Bit Lines, $\textbf{BL}_i$, and Source Lines, $\textbf{SL}_i$, we write into both MTJs in each memory cell, $\textbf{MTJ}_i$ and $\overline{\textbf{MTJ}_i}$, so that they hold complementary values. If $\textbf{MTJ}_i$ is in the $P$ state then $\overline{\textbf{MTJ}_i}$ will be in the $AP$ state and vice versa. This will result in a wide read margin during the read operation. 

After the termination of the write operation, in order to read the data stored in the MTJs, $\textbf{\textbf{RWL}}$ and $\overline{\textbf{RWL}}$ signals will be enabled, which results in activation of Read TGs of each memory cell, $\textbf{TGR}$. During the read operation, $\textbf{PR}$ and $\textbf{NR}$ transistors are turned on when $\textbf{RWL}$ and $\overline{\textbf{RWL}}$ are asserted, which provides the read path from $\textbf{VDD}$ to $\textbf{GND}$. The source of $\textbf{PR}$, which is a PMOS transistor, is connected to $\textbf{VDD}$ to provide strong one and the source of $\textbf{NR}$, which is an NMOS transistor, is connected to $\textbf{GND}$ to provide strong zero. A voltage divider circuit is designed as a result of resistance difference between the $\textbf{MTJ}_i$ and $\overline{\textbf{MTJ}_i}$, and the divided voltage can be observed at the $\textbf{D}_i$ nodes shown in Fig. \ref{subfig:fracclut} and Fig. \ref{subfig:fracsttclut}. According to the select tree input signals, shown as $\textbf{A}$, $\textbf{B}$, $\textbf{C}$, $\textbf{D}$, $\textbf{E}$, and $\textbf{F}$ in Fig. \ref{fig:lutfig}, using two inverters, the voltage on $\textbf{D}_i$ nodes will be amplified to generate the required output. Since the values stored in the $\textbf{MTJ}_i$ and $\overline{\textbf{MTJ}_i}$ devices are complementary, using one MTJ device to retain the data value and the other as the reference value will result in a wide read margin from $AP$ to $P$ \cite{Salehi2018BGIM:Applications}, which we leverage herein to increase the reliability of the read operation. 

In the proposed C-LUT design there is no need for an external clock or a large sense amplifier circuit. Furthermore, the proposed fracturable C-LUT can perform as a single $6$-input LUT or two $5$-input LUTs. The Operation mode of the proposed LUT is controlled using $\textbf{S5}$ and $\textbf{S6}$ signals. If $\textbf{S5}$ signal is enabled and $\textbf{S6}$ is disabled, then the C-LUT will be operating as two $5$-input LUTs and the outputs of the C-LUT will be $\textbf{OUT0}$ and $\textbf{OUT2}$. On the other hand, if $\textbf{S5}$ signal is disabled and $\textbf{S6}$ signal is enabled, then the C-LUT will be operating as a $6$-input LUT and $\textbf{OUT1}$ will be the C-LUT's output. The proposed fracturable C-LUT provides significantly higher functional flexibility at the expense of slightly more power consumption as studied in Section \ref{sec:sim-res}.

\section{Simulation Framework, Results, and Analysis}
\label{sec:sim-res}
\begin{figure}[!t]
\centering
\subfigure[]{
\includegraphics[width=3.0in]{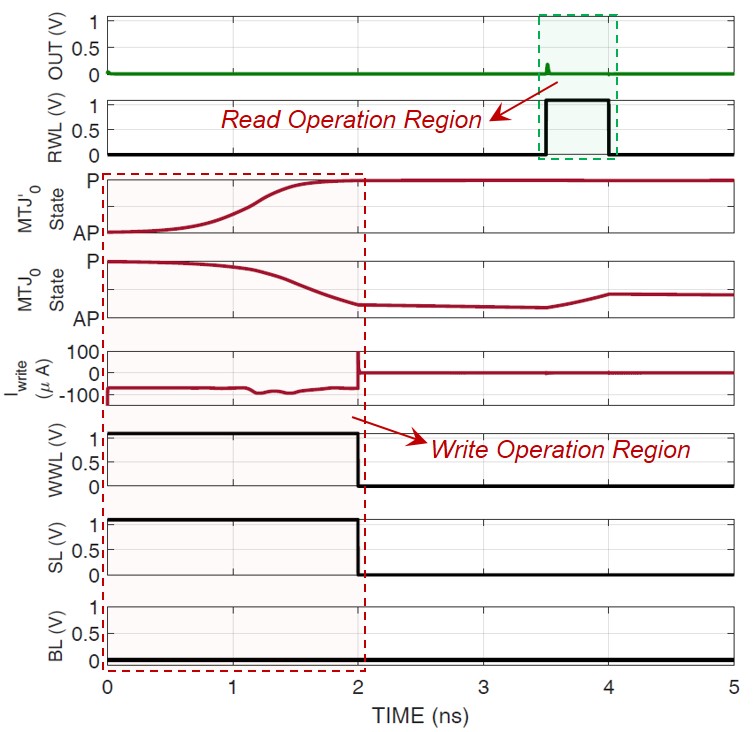}
\label{subfig:clut0}
}
\subfigure[]{
\includegraphics[width=3.0in]{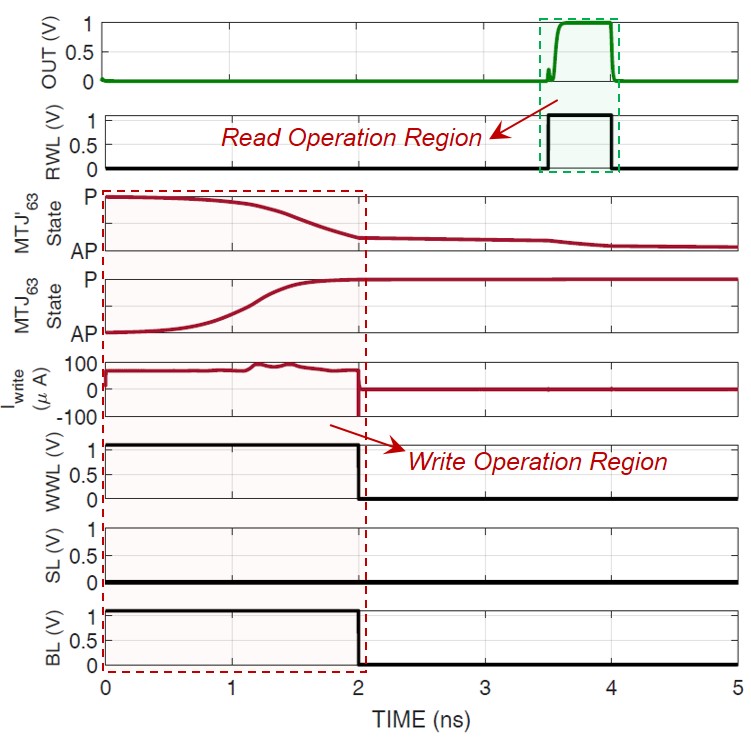}
\label{subfig:clut1}
}
%\vspace{-2mm}
\caption{Transient response of C-LUT implementing $6$-input OR operation for (a) $ABCDEF=``000000"$ input signal, and (b) $ABCDEF=``111111"$ input signal.}
%\vspace{-5mm}
\label{fig:clutfig}
\end{figure}

Herein, we use the HSPICE circuit simulator to validate the functionality of proposed C-LUT using $45$nm CMOS technology and the STT-MRAM model developed by Kim \textit{et al.} in \cite{Kim2015AStudies}. Figure \ref{subfig:clut0} and \ref{subfig:clut1} show the transient response of the C-LUT implementing a $6$-input OR operation for $ABCDEF=``000000"$ and $ABCDEF=``111111"$ input signals, respectively. In order to generate the current required for a write delay of less than $2$ns, the write transistors are required to be enlarged $4$-fold. As shown, the HSPICE simulations verify the correct functionality of our proposed C-LUT.

\begin{table}[!t]
%\small
\centering
\caption{Comparison between SRAM-LUT and MRAM-LUT.}
%\vspace{-5mm}
\label{tab:compare}
\setlength{\tabcolsep}{0.3em}
\begin{tabular}{llccccc}
\hline
\multicolumn{2}{l}{}      & \multicolumn{3}{c}{Power ($\mu W$)} 
															  & \multicolumn{2}{c}{Delay}
                                                                  				\\ \cline{3-7} 
\multicolumn{2}{l}{}                    & Read  & Write & Standby & Read  & Write\\ \hline
\multirow{3}{*}{SRAM LUT} & Logic ``0'' & 2.58  & 28.4  & 1.5     & 30 ps & 20 ps\\
                          & Logic ``1'' & 7.55  & 27.7  & 1.85    & 30 ps & 20 ps\\
                          & Average     & 5.06  & 25.08 & 1.67    & 30 ps & 20 ps\\\hline
\multirow{3}{*}{MRAM C-LUT}& Logic ``0'' & 14.38 & 81.16 & 0.31    & 20 ps & 2 ns\\
                          & Logic ``1'' & 19.91 & 81.25 & 0.31    & 60 ps & 2 ns\\
                          & Average     & 17.15 & 81.18 & 0.31    & 40 ps & 2 ns\\\hline       
\end{tabular}
\end{table}

\begin{table}[!t]
%\small
\centering
\begin{threeparttable}[b]
\caption{Area and Energy Consumption comparison between SRAM LUT and MRAM C-LUT.}
%\vspace{-5mm}
\label{tab:compare2}
\setlength{\tabcolsep}{0.1em}
\begin{tabular}{llcc}
\hline

		 & Features	     & SRAM LUT & MRAM C-LUT \\ \hline 
		 & Storage Cells & 384 MOS   & 128MTJ \\
Device 	 & Write/Control & 384 MOS  	& 256$\times$4 + 256 MOS\tnote{(1)} \\
Count  	 & Read     	 & 261 MOS  	& 267 MOS \\
         & Total   		 & 1029 MOS 	& 1547 MOS + 128 MTJ \\ \hline
Average Energy 	& Read   & 2.53 fJ  & 8.58 fJ \\
Consumption 	& Write  & 14 fJ    & 162.36 fJ \\ \hline
\end{tabular}
\begin{tablenotes}
\scriptsize
\item[(1)] Write transistors are 4$\times$ larger than minimum feature size.
\end{tablenotes}
\end{threeparttable}
\end{table}

% \iffalse
% \begin{table}[!t]
% %\small
% \centering
% \begin{threeparttable}[b]
% \caption{Area and Write Energy Consumption comparison between STT-MRAM C-LUT and SHE-MRAM C-LUT.}
% \label{tab:shemram}
% \setlength{\tabcolsep}{1em}
% \begin{tabular}{llcc}
% \hline
% 		&\multirow{2}{*}{Features} & \multicolumn{2}{c}{C-LUT} \\\cline{3-4}
%         &				& STT-MRAM 		  &	SHE-MRAM	   	\\\hline
%      	& Storage Cells & \multicolumn{2}{c}{128 MTJ}		\\
% Device  & Write/Control & \multicolumn{2}{c}{256 MOS + 128 MTJ\tnote{1}} \\
% Count   & Read          & \multicolumn{2}{c}{535 MOS + 6 MTJ}   \\
%         & Total         & \multicolumn{2}{c}{791 MOS + 128 MTJ}   \\\hline
% \multicolumn{2}{l}{Average Write Energy}  & \multirow{2}{*}{162.36 fJ}  & \multirow{2}{*}{53.59 fJ} \\
% \multicolumn{2}{l}{Consumption per Cell}  &			&	\\\hline
% \end{tabular}
% \begin{tablenotes}
% \scriptsize
% \item[1] Write circuit transistors for STT-MRAM C-LUT are $4\times$ larger than minimum feature size. Transistors with minimum feature size are used in the SHE-MRAM C-LUT.
% \end{tablenotes}
% \end{threeparttable}
% \end{table}
% \fi

\begin{table}[!t]
%\small
\centering
\begin{threeparttable}[b]
\caption{Iso-Delay Area and Write Energy Consumption comparison between STT-MRAM and SHE-MRAM C-LUTs.}
\label{tab:shemram}
\setlength{\tabcolsep}{0.4em}
\begin{tabular}{llcc}
\hline
&\multirow{2}{*}{Features} 		& \multicolumn{2}{c}{C-LUT} \\\cline{3-4}
    	    &					& STT-MRAM 		  					& SHE-MRAM \\ \hline
	    	& Storage Cells    	& 128MTJ               				& 128MTJ               \\
Device      & Write/Control    	& (256$\times$4)+256MOS \tnote{(1)} 	& 256+256MOS \tnote{(2)} \\
Count       & Read             	& 267MOS      						& 267MOS      \\
            & Total            	& 1547MOS+128MTJ    				& 779MOS+128MTJ     \\ \hline
\multicolumn{2}{l}{Average Write} & \multirow{2}{*}{162.3 fJ}             & \multirow{2}{*}{175.5 fJ}             \\
\multicolumn{2}{l}{Energy per Cell} & & \\\hline
\end{tabular}
\begin{tablenotes}
\scriptsize
\item[(1)] Write transistors are $4\times$ larger than minimum feature size.
\item[(2)] Write transistors with minimum feature size are used.
\end{tablenotes}
\end{threeparttable}
\end{table}

Table \ref{tab:compare} lists comparison results between the SRAM-LUT and proposed C-LUT in terms of power consumption and delay. The results show more than $80\%$ standby power reduction at the cost of increased write power which can be tolerated due to its infrequent occurrence of write operations in LUTs. There are three energy profiles in the FPGA LUT circuits: $(1)$ Read energy consumption during the FPGA normal operation, $(2)$ Standby energy for the LUTs that are not on the active datapath, which can constitute a significant portion of the FPGA fabric, and $(3)$ write energy that is consumed during the LUTs' configuration operation which occurs rarely. Table \ref{tab:compare2} provides an area and energy consumption comparison between SRAM-LUT and C-LUT. As listed, the structure of a $6$-input MRAM-based C-LUT requires $1,547$ MOS transistors plus $128$ MTJs, which can be fabricated on top of the CMOS transistors incurring low area overhead, while the conventional $6$-input SRAM-LUT includes $1,029$ MOS transistors. This results in an area overhead of roughly $50\%$ for C-LUT compared to SRAM-LUT, which is primarily induced by the write circuits. Thus, innovations are sought to reduce the area and energy consumption of the MRAM cell's write circuit to mitigate these issues. 
%%%%merge here
Recently, SHE-MRAM cells have attracted considerable attentions as an alternative for the conventional STT-MRAMs. Herein, we have used the SHE-MRAM device model proposed by Camsari \textit{et al.} \cite{Camsari2015ModularSpintronics} to realize a circuit-level simulation of our SHE-MRAM C-LUT. The results obtained exhibit that a TG-based write circuit with minimum-sized MOS transistors can produce the sufficient write current amplitude required for switching the SHE-MRAM's state in less than $2$ns. Thus, table \ref{tab:shemram} provides an iso-delay comparison between STT-MRAM and SHE-MRAM C-LUT in terms of device count and write energy. As listed, the SHE-MRAM C-LUT can achieve more than $49\%$ area reduction, while realizing comparable write energy consumption. Moreover, the SHE-MRAM C-LUT achieves at least $24\%$ device count reduction compared to SRAM-LUT.     

\begin{figure*}[!t]
\centering
\subfigure[ ]{
\includegraphics[width=2.2in]{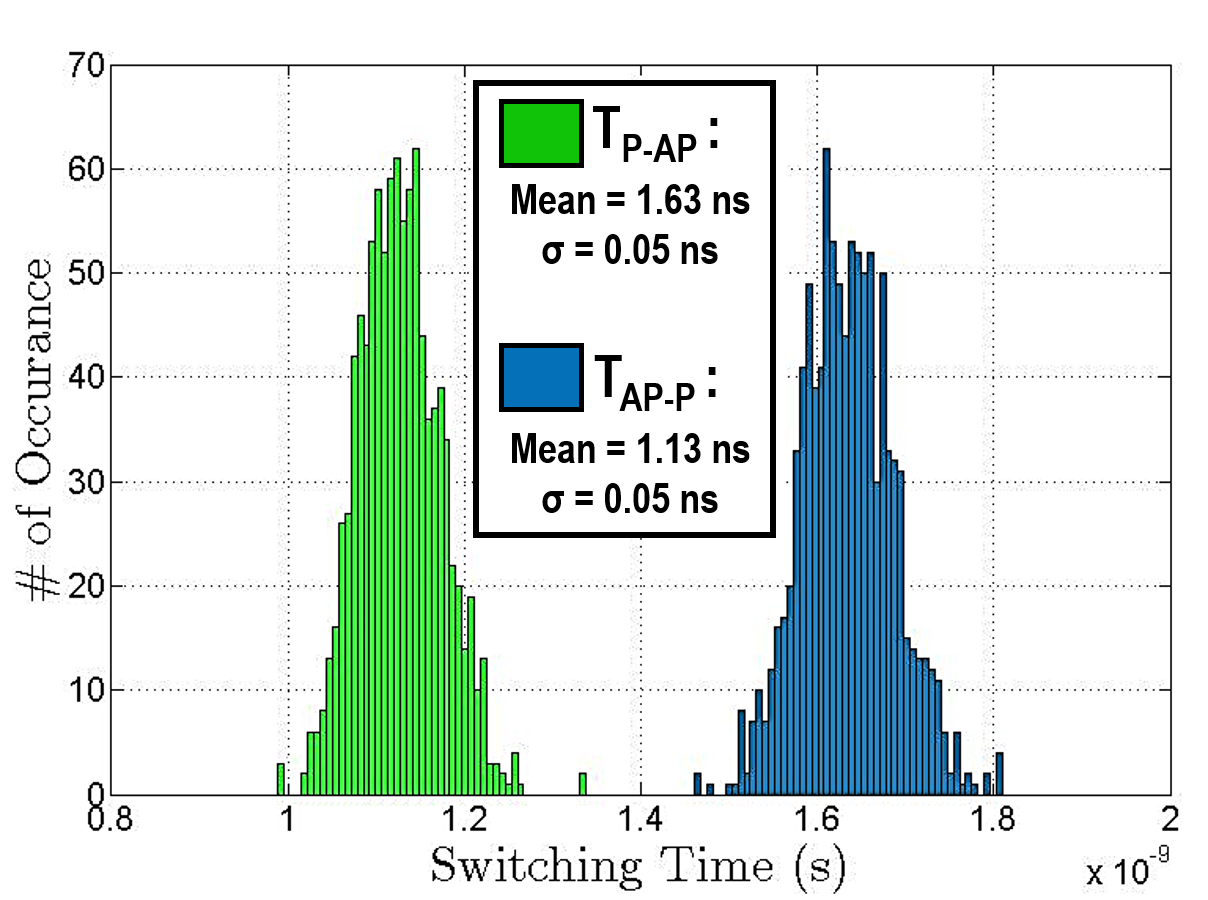}
\label{subfig:swt}
} %\vspace{-5mm}
\subfigure[ ]{
\includegraphics[width=2.2in]{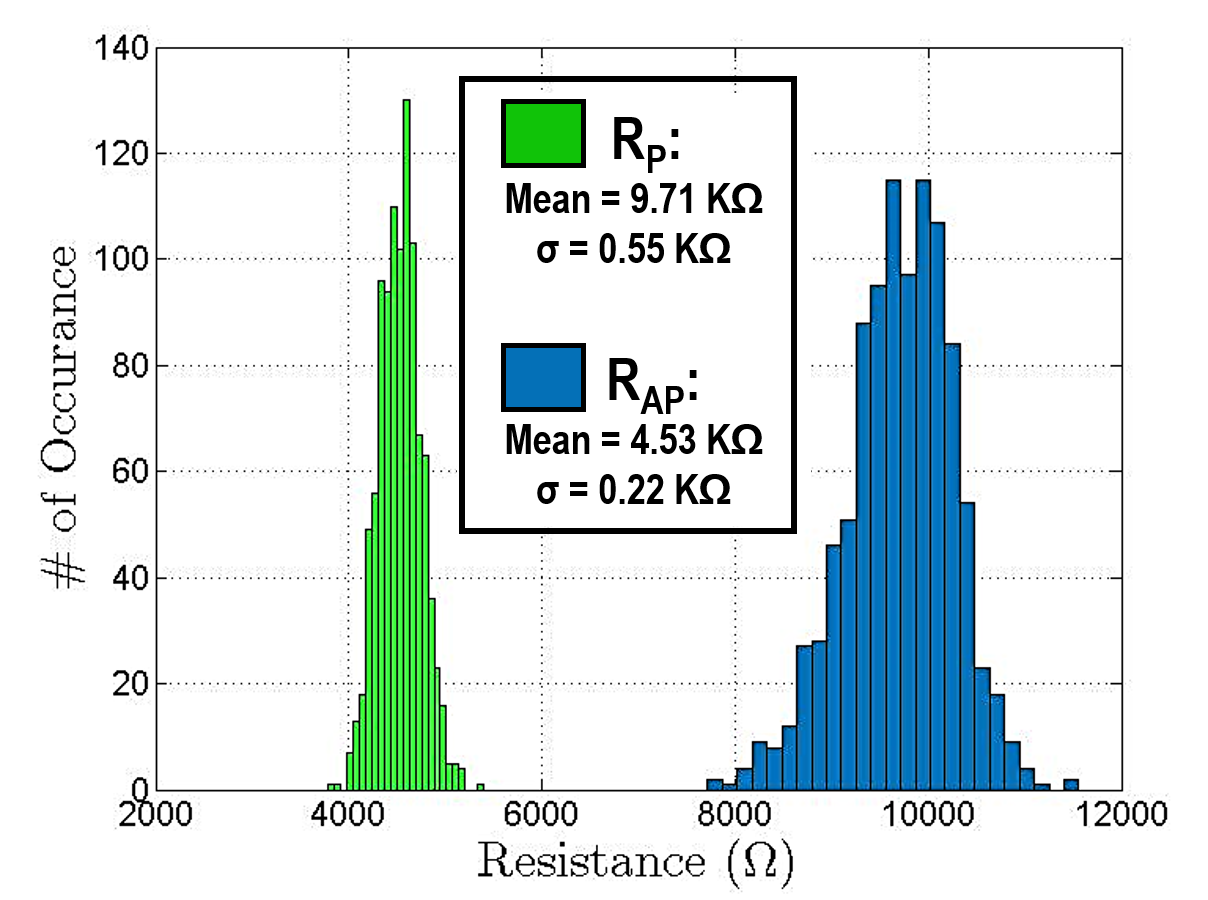}
\label{subfig:res}
} %\vspace{-5mm}
\subfigure[ ]{
\includegraphics[width=2.2in]{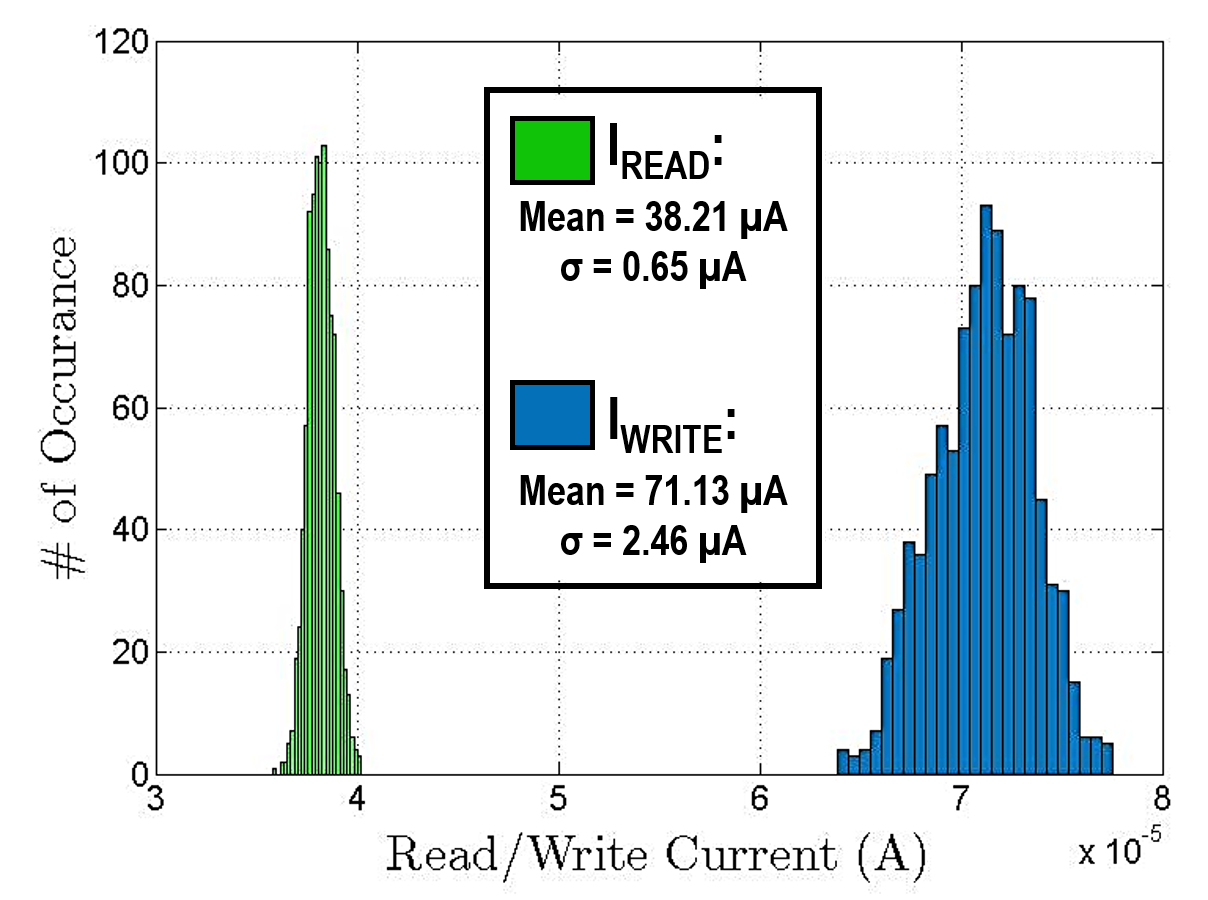}
\label{subfig:current}
} 
%\vspace{-2mm}
\caption{Simulation Results of $1,000$ MC instances for \subref{subfig:swt} $T_{P-AP}$ and $T_{AP-P}$ Switching Times, \subref{subfig:res} $R_{AP}$ and $R_{P}$ resistance states, and \subref{subfig:current} read, $I_{READ}$, and write, $I_{Write}$ currents.} 
\label{fig:reliability}
%\vspace{-1mm}
\end{figure*}

Furthermore, to analyze the reliability of the read and write operations of the proposed C-LUT, Monte Carlo (MC) simulation is performed to cover a wide range of PV scenarios that may occur in the fabricated device. The MC simulation is performed with $1,000$ instances considering the effects of PV on CMOS peripheral circuit and the MTJs. In particular, variation of $10\%$ for the MTJs' dimensions along with $10\%$ variation on the threshold voltage and $1\%$ variation on transistors dimentions are assessed. Fig. \ref{subfig:swt} depicts the distribution of the switching times for $T_{P-AP}$ and $T_{AP-P}$, Fig. \ref{subfig:res} illustrates the distribution of MTJ resistances in $R_{AP}$ and $R_{P}$  states, and Fig. \ref{subfig:current} shows the distribution of read, $I_{READ}$, and write, $I_{Write}$ currents for the $1,000$ MC instances. According to the MC simulation results, C-LUT provides reliable write performance resulting in less than $0.001\%$ write errors in $1,000$ error-free MC instances. In particular, results of the MC simulation show that the switching time for $P-AP$ is $1.63$ns on average and the switching time for $AP-P$ is $1.13$ns on average, which both fall under the $2$ns duration of the write operation, as depicted in Fig. \ref{subfig:swt}. Additionally, since the states of the MTJs are differential, they provide a wide read margin and as a result there are less than $0.001\%$ read errors caused by PV based on the $1,000$ error-free MC simulation results. Furthermore, our proposed C-LUT does not suffer from read disturbance due to the small read current compared to the write current as shown in Fig. \ref{subfig:current}. According to our MC simulation results, the read current is $38.21\mu$A on average, which is significantly lower than the write current that is $71.13\mu$A on average. 
%Thus, according to (\ref{eq:avgrm}), our proposed C-LUT maintains a significantly wide RM.

\section{Conclusion}
\label{sec:conclusion}
To overcome the conventional SRAM-LUT limitations such as high static power, volatility, and low logic density, we have proposed a novel LUT design using spin-based devices. The proposed C-LUT is a clockless design and a suitable candidate for combinational logic, which can also be combined with a flip-flop circuit to implement sequential logic. According to our simulation results, the standby power dissipation of the proposed C-LUT is $0.31\mu$W, which is reduced by $5.4$-fold compared to the SRAM-based LUT. Moreover, the structure of the proposed SHE-MRAM based C-LUT includes $250$ and $768$ fewer transistors compared to the SRAM-based LUT and the STT-MRAM based C-LUT, respectively. Additionally, according to the process variation reliability analysis, the C-LUT circuit exhibits $< 0.001\%$ error rate for read and write operations in presence of variations spanning both transistors and MTJs. 

% In an attempt to overcome the conventional SRAM-LUT limitations such as high static power, volatility, and low logic density, we have proposed a novel LUT design using spin-based devices called C-LUT. The proposed C-LUT is a clockless design and a suitable candidate for combinational logic, which can also be combined with a flip-flop circuit to implement sequential logic. According to our simulation results, the standby power dissipation of the proposed C-LUT is only $0.31\mu$W, which is reduced by $5.4$-fold compared to the SRAM-based LUT. Moreover, the structure of the proposed SHE-MRAM C-LUT includes $250$ and $768$ fewer transistors compared to the SRAM-based LUT and the STT-MRAM C-LUT, respectively. Additionally, according to the process variation reliability analysis, the C-LUT circuit exhibits less than $0.001\%$ error rate for read and write operations in presence of variations spanning both transistors and MTJs. 
\section*{Acknowledgement}
This work was supported in part by the National Science Foundation (NSF) through ECCS-$1810256$.
\balance

\bibliographystyle{ACM-Reference-Format}
%\bibliography{soheil.bib,ramtin.bib}
\bibliography{soheil}
\end{document}